# Computational Investigation of the Magnetization Reversal and Magnetoresistive Behavior of Nanoscale Spin Valve Elements

*Swapnil Barman*[1,*]

[1]Technical Research Centre, S. N. Bose National Centre for Basic Sciences, JD Block, Sector III, Salt Lake City, Kolkata – 700106, India

[*]*E-mail:* swapnil.barmanamity@gmail.com

**Abstract**

Investigation of the magnetic switching and magnetoresistive behaviour of nanoscale spin valve elements (SVs) of varying physical parameters such as shape, element size, dimensional aspect ratio, and array size is of vital importance for their application in future magnetic memory and storage devices. We have inspected the magnetic switching mechanism and magnetoresistive behaviour of nanoscale SV elements (Co/Cu/$Ni_{80}Fe_{20}$) and arrays of these elements, with each layer having a thickness of 10 nm and rectangular and elliptical shapes with varying lateral aspect ratios (ARs) and varying interelement spacing for the arrays, by finite difference method-based micromagnetic simulation. We observe that the elements with higher AR show the $Ni_{80}Fe_{20}$ and Co layers forming antiparallel states in the plateau, similar to synthetic antiferromagnets. For lower AR, more complex quasi-uniform magnetic microstates are observed, which are even more intricate for elliptical elements. The elliptical elements with an AR of 1.25 show coherent and predictable magnetic switching behaviour, demonstrating their appropriateness for integration into magnetic memory devices. We observe a gradual increase in magnetoresistance (MR%) with the increase in AR and the decrease in interelement spacing. The magnetic flux density experiences a reduction with an increase in the inter-elemental spacing.

## 1. Introduction

Charge-based electronics are based on transistors. Hence, they are facing problems in scalability and thermal runaway problems. In present microelectronic devices the commonly used memories are static RAM (SRAM) [1-2], dynamic RAM (DRAM) [3-4] and flash memory [5]. These are all capacitive technologies which store memory in charge states. As times have progressed, these memory devices have been scaled down efficiently to reach higher speeds and increased density of memory chips in a cost-effective way [6]. However, charge based storage devices are reaching the physical limits of scalability. The emergence of zero capacitance memory (ZRAM) [7], advanced RAM (ARAM), zero impact ionization memory (Z2RAM) have provided impetus in the charge-based memory devices. However, designing and developing universal memory is the next step in the progression of memory. Future universal memory [8] is expected to combine the high density of DRAM, non-volatility of flash memory and high speed of SRAM. Therefore, development of new concepts of memories based on a different storage principle will be important.

As a replacement for charge-based RAM, resistive memories such as resistive random-access memory (RRAM) [9], phase change memory [10] and magnetoresistive random-access memory (MRAM) [11] hold a lot of promise. MRAM is one of the early success stories in the field of magnetoelectronics, also known as spintronics, which employs the electron's spin degree of freedom in addition to its charge degree of freedom [12]. Unlike charge-based RAM, data in MRAM is processed using magnetic elements rather than electric charge or current. The elements are made up of two ferromagnetic thin layers separated by a thin insulating layer, each of which may have a magnetization. To store memory, the magnetic state of one of those two ferromagnetic layers is fixed in one direction (reference layer: RL), while the magnetization of the other layer is free to change (free layer: FL). The FL can switch between two magnetic states, parallel and anti-parallel to the RL's fixed magnetization direction. The antiparallel and parallel states have different magnetoresistance values, which are mapped to the binary memory states ‘0' and ‘1'. This device is referred to as a magnetic tunnel junction (MTJ) [13], and it is the most basic unit for an MRAM bit. A memory device is made up of a network of these memory cells.

The giant magnetoresistance (GMR) and tunnel magnetoresistance (TMR) effects are the two most common magnetoresistance phenomena. Julliere discovered TMR in a Fe/Ge/Co junction at low temperature (T 4.2K) in 1975 [14]. It was discovered in an MTJ, a pillar made up of two ferromagnetic layers separated by a thin insulating layer. The pillar's resistance is determined by the magnetization orientation of the layers in relation to one another and is expressed as:

$$TMR = \frac{R_{AP} - R_P}{R_P} \quad [1]$$

where, $R_{AP}$ and $R_P$ correspond to the resistance in antiparallel (high resistance) and parallel (low resistance) magnetic states. The following can be deduced from this. If free states with the same spin orientation are visible, electrons with a particular spin orientation (‘spin-up' or ‘spin-down') can tunnel from one ferromagnetic layer to another through a non-conducting thin insulating barrier layer. The majority spin ('spin-up') and minority spin ('spin-down') electrons will easily tunnel through the barrier to the other ferromagnetic layer, filling the majority ('up') and minority ('down') states, resulting in a low resistive state in the parallel state. The majority spin (‘spin-down') and minority spin (‘spin-up') electrons from the first ferromagnetic layer, on the other hand, will fill the minority (‘down') and majority (‘up') states in the second ferromagnetic layer, respectively, resulting in a low conductance, i.e. a high resistive state, in the antiparallel state. GMR occurs in a standard spin valve device, which consists of two thin ferromagnetic layers separated by a thin metallic spacer. The magnetoresistance (MR) is caused by spin-dependent electron scattering at the interface, which is influenced by the parallel and antiparallel magnetic states of the two ferromagnetic layers.

As a result, the magnetic switching behavior of the magnetic layers in MRAM is critical. Data writing and reading in traditional MRAM with cross-wire architecture is accomplished by sending electrical current through the wires (write and read lines), with the Oersted magnetic field produced by the current achieving magnetic switching or reversal of the layers. With the miniaturisation of MRAM cells, the necessary switching field and current become extremely high, resulting in energy loss due to Joule heating and a subsequent thermal runaway problem. After that, a new technology called spin transfer torque MRAM (STT-MRAM) was discovered to solve the problem [15]. A spin-polarized current is used to switch the electrons' magnetization [16-17]. This effect occurs in an MTJ or a spin-valve, and STT tunnel junctions are used in STT-MRAM systems (STT-MTJ). As a result, Everspin has begun manufacturing 1GB STT-MRAM chips, which IBM will use in its next-generation FlashCore modules [18].

Here, computational micromagnetic simulation was used to investigate the magnetic switching and magnetoresistive behavior of nanoscale spin valve elements in the form of $Co/Cu/Ni_{80}Fe_{20}$ (Permalloy, Py hereafter). The spin valve components were modelled in two different shapes, rectangular and elliptical, with the lateral aspect ratio (eccentricity) of the cells varied. After that, we looked at elliptical element arrays. With shifts in the aspect ratio for both the shapes and the cell spacing for the arrays of components, a drastic shift in magnetic hysteresis loops (magnetization vs. applied magnetic field), saturation field, coercive field, remanence, and switching behavior is observed. To better understand the observed behavior, we simulated the magnetization reversal states.

## 2. Methods

The LLG micromagnetic simulator [19] was used to simulate magnetic hysteresis loops (magnetization ($M$) vs. external magnetic field ($H$)). The Landau-Lifshitz-Gilbert (LLG) equation [20] given below governs the magnetization dynamics in this case.

$$\frac{d\boldsymbol{M}(t)}{dt} = -\gamma\left[\boldsymbol{M}(t) \times \boldsymbol{H}_{eff}(t)\right] + \frac{\alpha}{M_s}\left[\boldsymbol{M}(t) \times \frac{d\boldsymbol{M}(t)}{dt}\right] \quad [2]$$

Here, the first term represents the magnetization precession torque under the application of a magnetic field ($\boldsymbol{H}$) while the second term represents the damping torque. $\gamma$ is the gyromagnetic ratio, $\boldsymbol{M}$ is the magnetization vector and $\boldsymbol{H}_{eff}$ is the effective magnetic field consisting of various terms as given below.

$$\boldsymbol{H}_{eff} = \boldsymbol{H}_{ext} + \boldsymbol{H}_{demag} + \boldsymbol{H}_{ani} + \boldsymbol{H}_{exc} \quad [3]$$

The exchange interaction field is $H_{ex}$, the demagnetizing field is $H_d$, and the anisotropy field is $H_K$. The exchange energies of both the Py and the Co layers are included in $H_{ex}$ in the Co/Cu/Py trilayer portion. $H_d$ takes into account the demagnetizing fields of both magnetic layers, and it is affected by the length-scale, shape, and aspect ratio of the components. It is expected to change when we change the above-mentioned parameters in our simulation. The magneto-crystalline anisotropy of the two layers makes up $H_K$. $M_s$ is the saturation magnetization, and $\alpha$ is the dimensionless damping coefficient. The LLG equation is solved using the finite difference method in the simulation. The geometry of the investigated spin-valve elements is divided into cuboidal cells with dimensions of 10 nm × 10 nm × 10 nm. The software's graphical user interface (GUI) is used to describe the element geometry, dimensions, number of layers, and properties. Two sample shapes, rectangular and elliptical, were used in these

**Table 1.** Simulation parameters are listed in the table for two magnetic layers.

| **Material** | **$M_s$ (emu/cm$^3$)** | **$\gamma$ (MHz/Oe)** | **$A_{ex}$ (μerg/cm)** | **$K_{u2}$ (erg/cm$^3$)** |
|---|---|---|---|---|
| **Py ($Ni_{80}Fe_{20}$)** | 800 | 17.60 | 1.050 | 1000.00 |
| **Co** | 1414 | 17.60 | 3.050 | 400000.00 |

simulations. with the length (L) of the components varied as 250, 200, 150, 125, 100, and 50 nm and the width set at 100 nm. Consequently, the lateral aspect ratio (AR) varies as 2.5, 2.0, 1.5, 1.25, 1.0 and 0.5. In addition, we simulated 2×2 and 3×3 elliptical element arrays with AR

= 1.25. The cell spacing was chosen as 10, 25, 50, 100, 250, and 500 nm. Each layer's thickness is set to 10 nm. Table 1 lists the material parameters that were used in the simulation. Each layer's exchange constant is $A_{ex}$, the uniaxial magnetic anisotropy constant is $K_{u2}$, and the interlayer exchange is $A_{ij}$, which was set as zero, in order to understand the reversal mechanism originated from purely dipolar interactions between the layers.

LLG equation is an ordinary differential equation which is solved using 'Time Integration – Rotation Matrices' method with a three-dimensional (3D) complex fast Fourier transformation (FFT) method. In the simulation, 'Time Step' of 1 ps, maximum iteration of 25000 with damping coefficient $\alpha = 1.0$ was used for fast convergence at each step to find the equilibrium magnetization at each magnetic field step. For calculating the hysteresis loops the magnetic field (H) was ramped between up to ±5000 Oe depending on the saturation field of the sample. The hysteresis loops and magnetoresistance loops averaged over the sample volume are collected for each sample. In addition, the magnetization distributions over the magnetic layers for each sample have been studied to understand the reversal modes and spatial coherence of magnetization switching.

## 3. Results and Discussions

### 3A. Effects of Element Shape on Magnetic Switching and Magnetoresistive Behavior of Spin Valve Elements

The simulated magnetic hysteresis loops with magnetic field applied along x-axis, showing the magnetization switching behavior, magnetoresistance and extracted magnetic parameters ($H_c$, $M_r$ and $H_{sat}$) for the rectangular Co/Cu/Py spin valve elements with varying AR are shown in Fig. 1. For AR = 0.5, a hard-axis-like loop is observed with a small hysteresis at the center.

The simulated magnetic hysteresis loops with magnetic field applied along x-axis, showing the magnetization switching behavior, magnetoresistance and extracted magnetic parameters ($H_c$, $M_r$ and $H_{sat}$) for the rectangular Co/Cu/Py spin valve elements with varying AR are shown in Fig. 1. For AR = 0.5, a hard-axis-like loop is observed with a small hysteresis at the center. This is because the elements undergo a transition from easy-axis to hard-axis of its shape anisotropy *w.r.t.* the applied field direction with the reduction of AR. For 1.0 ≤ AR ≤ 2.5, we observe two-step switching, with the squareness of the loops increasing, with the increase in AR. The size of the plateaus also increases, as we increase the AR. We observe the systematic decrease in $H_{sat}$ until 1.5 kOe at AR = 1.5, after which it decreases gradually and the systematic increase in $H_c$ with the increase in AR. For AR = 0.5, a narrow MR loop is observed with large spacing. For 1.0 ≤ AR ≤ 2.5, we observe that the squareness of the MR loops increases while they become wider with the increase in AR, and the space in between decreases. The loops almost merge at AR = 2.5. We observe the gradual increase in MR % to about 12% with the increase in AR and as the loops become wider.

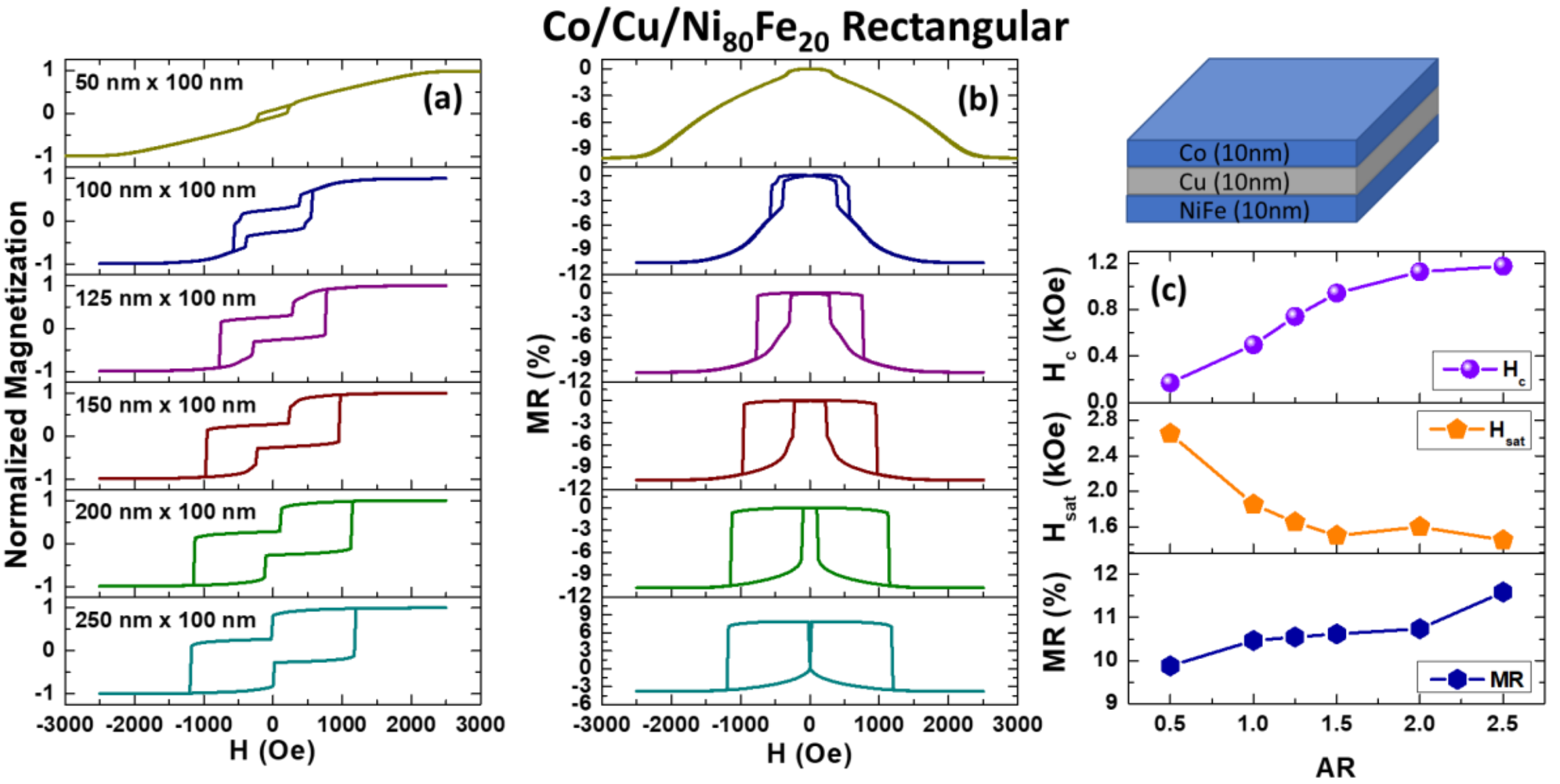

Fig. 1. Simulated magnetic hysteresis loops (normalized magnetization vs. magnetic field), magnetoresistance and extracted magnetic parameters ($H_c$, $M_r$ and $H_{sat}$) as a function of AR for rectangular Co/Cu/Py spin valve elements with varying AR. The maximum magnetic field was adjusted according to the saturation of magnetization for each element. The magnetic field is applied along x-axis.

In order to study the effects of element shape, we have simulated elliptical elements with dimensions as described before. The simulated magnetic hysteresis loops with magnetic field applied along x-axis, showing the magnetization switching behavior, magnetoresistance and extracted magnetic parameters ($H_c$, $M_r$ and $H_{sat}$) for the elliptical Co/Cu/Py spin valve elements with varying AR are shown in Fig. 2. For the AR = 0.5, we observe a hard-axis-like loop with no hysteresis at the center of the loop. For the AR = 1.0, we observe a single-step switching with increased squareness of the loop than AR = 0.5 with some hysteresis at the center. For 0.5

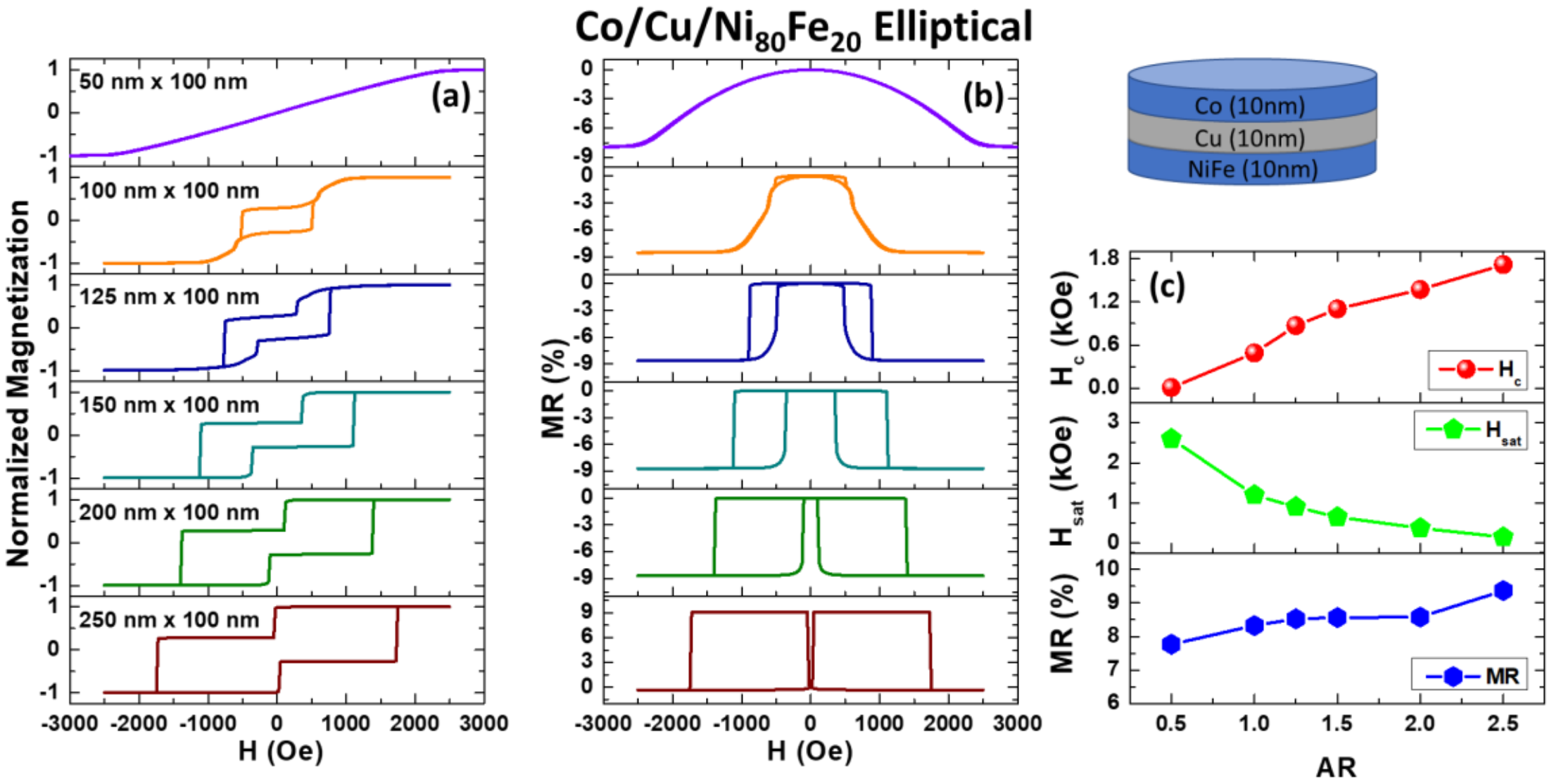

Fig. 2. Simulated magnetic hysteresis loops (normalized magnetization vs. magnetic field), magnetoresistance and extracted magnetic parameters ($H_c$, $M_r$ and $H_{sat}$) as a function of AR for elliptical Co/Cu/Py spin valve elements with varying AR. The maximum magnetic field was adjusted according to the saturation of magnetization for each element. The magnetic field is applied along x-axis.

$\leq$ AR $\leq$ 1.0, the switching observed in incoherent. For 1.25 $\leq$ AR $\leq$ 2.5, we observe a two-step switching, with the squareness of the loops increasing further, with the increase in AR and the magnetic field switching observed is coherent. The size of the plateaus also increases, as we increase the AR. We observe a systematic decrease in $H_{sat}$ and increase in $H_c$ with the increase in AR. For AR = 0.5, a narrow MR loop is observed with large spacing between the two loops. For 1.0 $\leq$ AR $\leq$ 2.5, we observe a systematic increase in the squareness of the MR loops while they also become wider with the increase in AR, while the spacing between the two loops decreases. The loops almost merge at AR = 2.5. Here too, we observe a gradual increase in MR % up to > 9% with the increase in AR and as the loops become wider.

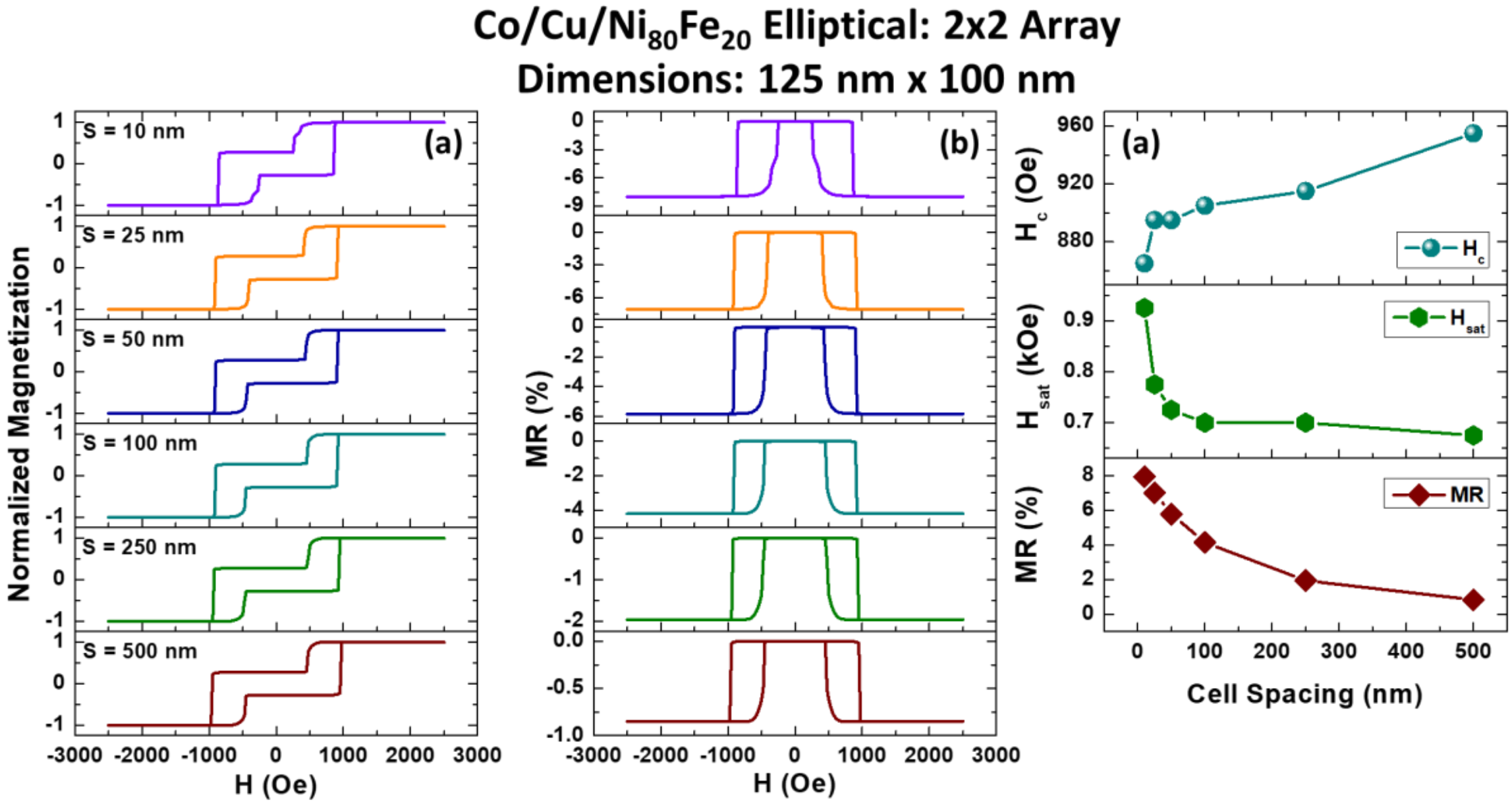

Fig. 3. Simulated magnetic hysteresis loops (normalized magnetization vs. magnetic field), magnetoresistance and extracted magnetic parameters ($H_c$, $M_r$ and $H_{sat}$) as a function of AR for 2×2 elliptical arrays of Co/Cu/Py spin valve elements with varying cell spacing. The maximum magnetic field was adjusted according to the saturation of magnetization for each element. The magnetic field is applied along x-axis.

We further studied the effects of dipolar interactions in an array of spin valve elements on their magnetization reversal and MR behavior. This is crucial as complex the inter-element dipolar interaction between the elements can significantly modify the magnetic switching field and MR as well as the spatial coherence of switching. A cross-talk between the elements is not desirable for their independent operation, whereas elements cannot be placed too far away to avoid the cross-talk for the benefit of areal density of the magnetic recording technology. The simulated magnetic hysteresis loops with magnetic field applied along x-axis, showing the magnetization switching behavior, magnetoresistance and extracted magnetic parameters ($H_c$, $M_r$ and $H_{sat}$) for the 2×2 elliptical arrays of Co/Cu/Py spin valve elements with varying AR is shown in Fig. 3. We took into consideration, 2×2 elliptical arrays with cell spacings of 10 nm ≤ S ≤ 500 nm. For all the cell spacings, two-step switching is observed and the squareness of the loops increases with the increase with the cell spacing. The size of the plateaus increases as we increase the cell spacing. We observe a sharp decrease in $H_{sat}$ from S = 10 nm to S = 25 nm, and a gradual decrease beyond that. We also observe a sharp increase in $H_c$ from S = 10 nm to S = 25 nm, and a gradual increase after that. For $0.5 \leq S \leq 2.5$, we observe that the squareness of the MR loops increases while they become narrower with the increase in cell spacing, and the space between increases gradually. We observe a gradual decrease in MR % from about 8% to about 2% with the increase in cell spacing and as the loops become narrower.

When we increased the number of elements in the array, the behavior remained qualitatively similar despite small quantitative changes. The simulated magnetic hysteresis loops with magnetic field applied along x-axis, showing the magnetization switching behavior, magnetoresistance and extracted magnetic parameters ($H_c$, $M_r$ and $H_{sat}$) for the 3×3 elliptical arrays of Co/Cu/Py spin valve elements with varying AR is shown in Fig. 4. We took into consideration, 3×3 arrays with cell spacings of 10 nm ≤ S ≤ 250 nm. For cell spacings of 10 nm ≤ S ≤ 50 nm multi-step switching is observed. For cell spacings of 150 nm ≤ S ≤ 250 nm two-step coherent switching is observed with the squareness of the loops increasing with the increase in cell spacing. The size of the plateaus increases as we increase the cell spacing. We observe the large decrease in $H_{sat}$ from ~850 Oe at S = 10 nm to ~650 Oe at S = 50 nm, after which it remains constant at 650 Oe and large increase in $H_c$ from S = 10 nm (~815 Oe) to S = 25 nm (~880 Oe), after which it increases systematically and slows down at S = 150 nm. For $0.5 \leq S \leq 2.5$, we observe that the squareness of the MR loops increases while they become narrower with the increase in cell spacing, and the space between increases gradually. We observe the gradual decrease in MR % with the increase in cell spacing and as the loops become narrower.

### 3B. Effect of Element Shape on the Magnetization Reversal Modes of the Spin Valve Elements.

Understanding the magnetization reversal mechanism is important in nanoscale magnetism and its application in memory devices [21]. We have further simulated the magnetization distributions (maps) of two magnetic layers during the magnetization reversal (switching). In layer magnetization partially retains towards +x direction although its magnetization at the central parts starts to deviate from the +x direction forming an opposite S-state *w.r.t.* the Py layer. At $H$ = -250 Oe, where a minor switching in the hysteresis loop is observed (4), the Co layer switches to -x direction but the Py layer switches back to the +x direction. This unusual observation occurs due to the fact that at this small external magnetic field the dynamics is dominated by the dipolar interaction field, the latter favors antiparallel alignment of the neighbouring magnetic layers. At the -ve saturation field (6) both the layers coherently switch their magnetization to -x-direction. For AR = 1.5, in the positive saturation (1), both the Co and Py layers are fully aligned along +x-direction at +ve saturation field. The edge regions of the Co layer are deviated from +x-direction, and the Py layer forms an S-state as we descend to (2). At the 1$^{st}$ plateau (3) the Py layer switches coherently, while Co layer retains its magnetization direction and gets more demagnetized at the end of the 1$^{st}$ plateau (4). At the -ve saturation (3) both the layers switch towards the -ve field direction. For AR = 2.5, in the positive saturation (1), both the Co and Py layers are fully aligned along +x-direction at +ve saturation field. Their edge regions are deviated from +x-direction as we descend to (2). At the 1$^{st}$ plateau (3) the Py layer switches coherently, while Co layer retains its magnetization direction and gets more demagnetized at the end of the 1$^{st}$ plateau (4). At the -ve saturation (3) both the layers switch towards the -ve field direction.

Fig. 5. we show the magnetization maps of the Co and the Py layers at some important magnetic field values for three rectangular shaped spin valve elements with AR of 2.5, 1.5 and 0.5. The magnetization maps are placed next to the corresponding hysteresis loops where the arrows indicate the field values at which the maps are simulated. For AR = 0.5, both Co and Py layers are fully aligned along the +x direction at the +ve saturation field (1). While descending down the loop, the Co and Py layers partially get demagnetized at (2). At $H$ = -200 Oe (3) the Py layer partially switches its magnetization towards -x direction forming an S-state, while the Co

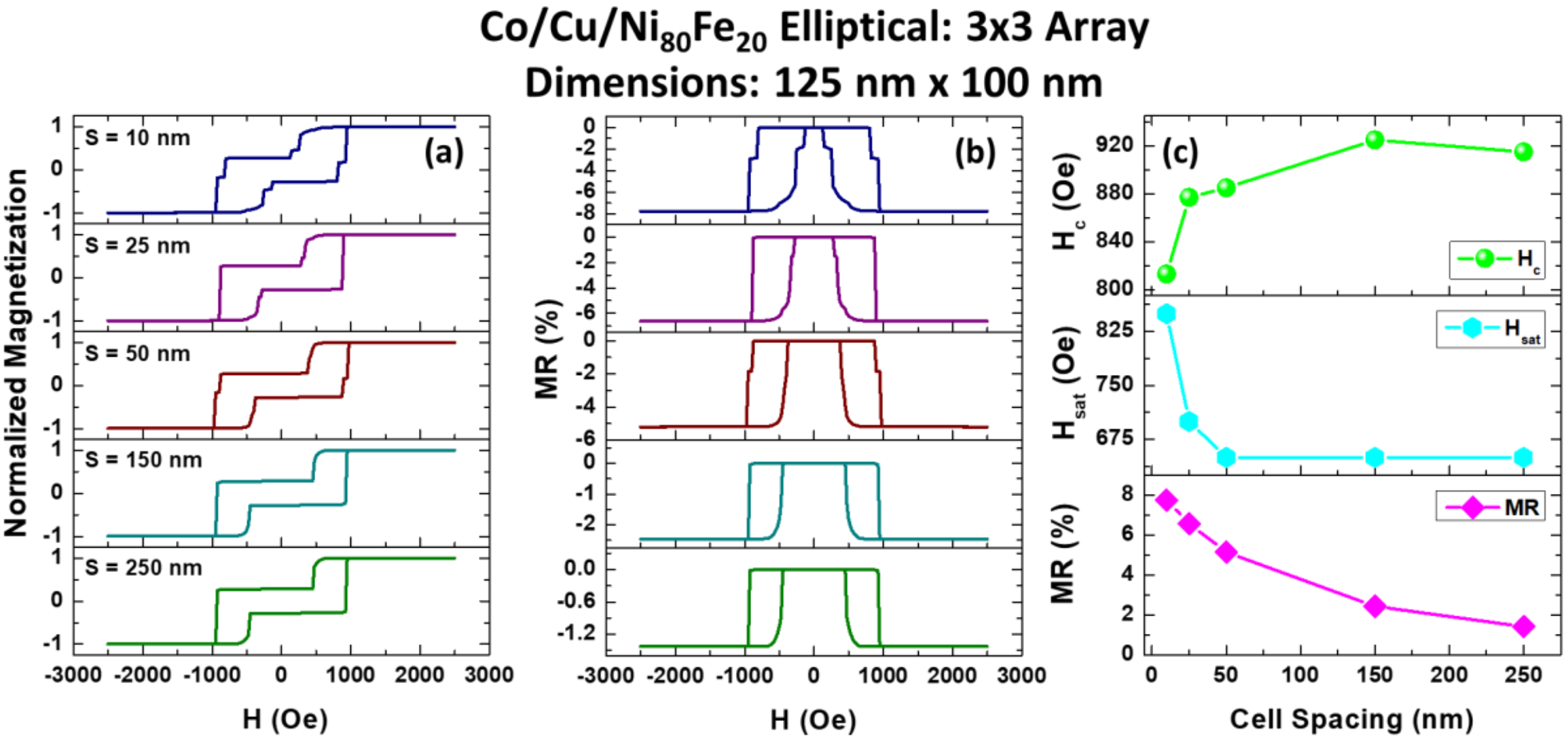

Fig. 4. Simulated magnetic hysteresis loops (normalized magnetization vs. magnetic field), magnetoresistance and extracted magnetic parameters ($H_c$, $M_r$ and $H_{sat}$) as a function of AR for 3×3 elliptical arrays of Co/Cu/Py spin valve elements with varying cell spacing. The maximum magnetic field was adjusted according to the saturation of magnetization for each element. The magnetic field is applied along x-axis.

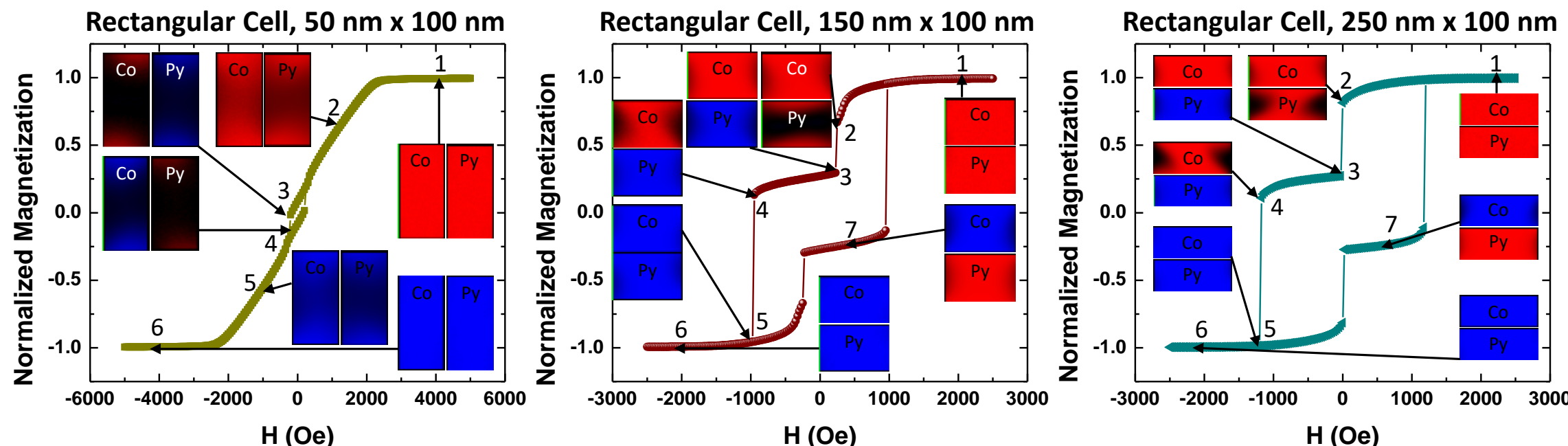

Fig. 5. Magnetization reversal states of Co and Py layers of the rectangular spin valve elements of three different aspect ratios. The magnetic fields corresponding to which the magnetization maps are presented are marked by numbers.

In Fig. 6. we show the magnetization maps of the Co and the Py layers at some important magnetic field values for four different elliptical shaped spin valve elements with AR of 2.5, 1.25, 1.0 and 0.5. The magnetization maps are placed next to the corresponding hysteresis loops where the arrows indicate the field values at which the maps are simulated. For AR = 0.5, at the +ve saturation field (1) the magnetization of Co and Py layers are aligned along the +x direction. At H = 0 Oe (2), both the Py and Co layers enter into two opposite edge domain states. The magnetic states change continuously and at the -ve saturation field (3) both the layers fully switch their magnetization towards -x direction. For AR = 1.0, at the +ve saturation field (1), the magnetization of the Co and Py layers are aligned along the +x direction. While approaching the 1st plateau (2), the Py layer enters into an S-state while Co layer retains its magnetization direction. At the end of the 1st plateau, the Py layer completely switches to –ve magnetization (3). After the switching to (4), the Co layer switches to -x direction but the Py layer switches back to the +x direction due to strong dipolar interaction between the two layers. At the -ve saturation (5) both the layers fully switch their magnetization towards -x direction. For AR = 1.25, at the +ve saturation field (1), the magnetization of the Co and Py layers are aligned along the +x direction. Till the end of the 1st plateau (4), the Co layer retains its original magnetization, while the Py layer coherently switches its magnetization to -ve direction at (2). At the -ve saturation (5) both the layers fully switch their magnetization towards -x direction. For AR = 2.5, clear two-step switching is observed. Here, at the +ve saturation field (1), the magnetization of the Co and Py layers are aligned along the +x direction as usual. At the 1st plateau (2), the Co layer retains its original magnetization, while the Py layer coherently switches to –ve magnetization. At the end of this plateau the Co layer also coherently switches its magnetization and at the -ve saturation field (3) both the layers have their magnetization switched towards -x direction. The shape anisotropy of this sample along the applied field direction helps this stepwise coherent switching of magnetization. To reveal the role of inter-element dipolar coupling on the magnetization reversal mechanisms of these spin valve elements, we investigate the magnetization maps of the Co and the Py layers at some important magnetic field values for 2×2 elliptical arrays of spin valve elements with AR of 1.25 and two different cell spacings. In Fig. 7. the magnetization maps are placed next to the corresponding hysteresis loops where the numbers indicate the field values at which the maps are simulated. For cell spacing S = 10 nm, at the +ve saturation field (1), both the Co and Py layers are aligned along the +x direction. At point (2), we notice that the Py elements starts to demagnetize at their corners while the Co elements maintain their magnetization direction. At points (3, 4), the Py forms C-states. At point (5), the elements in the Py layer completely switch their magnetization direction to -x. At point (6), switching of the Co elements nucleate at their edges and at the second switching point (7), the elements in the Co Layer also switch their magnetization direction to -x direction. At a large cell spacing (S = 250 nm) where the inter-element interaction field become much weaker the reversal behavior does not change significantly. The only discernible change is observed in the first switching step, where the Py elements initially formed S-like state (2) followed the complete switching of the top row. elements, while the bottom row elements still retain the S-like state (3). From the 1st plateau (4) onward, the reversal behavior for this sample is identical to what has been observed for the sample with S = 10 nm. This indicated that once the initial reversal barrier is crossed the in-element interaction field.

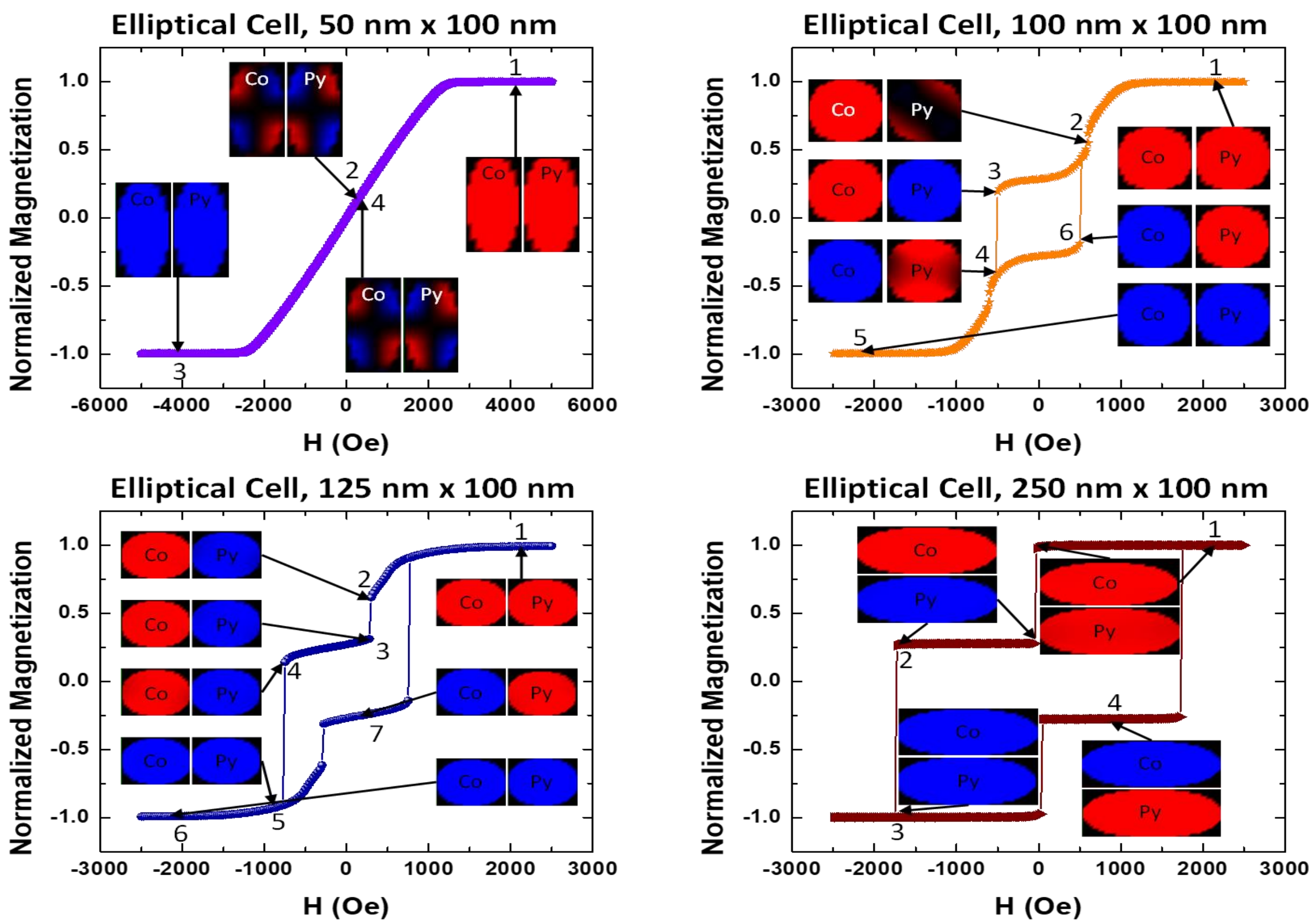

Fig. 6. Magnetization reversal states of Co and Py layers of the elliptical spin valve elements of four different aspect ratios. The magnetic fields corresponding to which the magnetization maps are presented are marked by numbers.

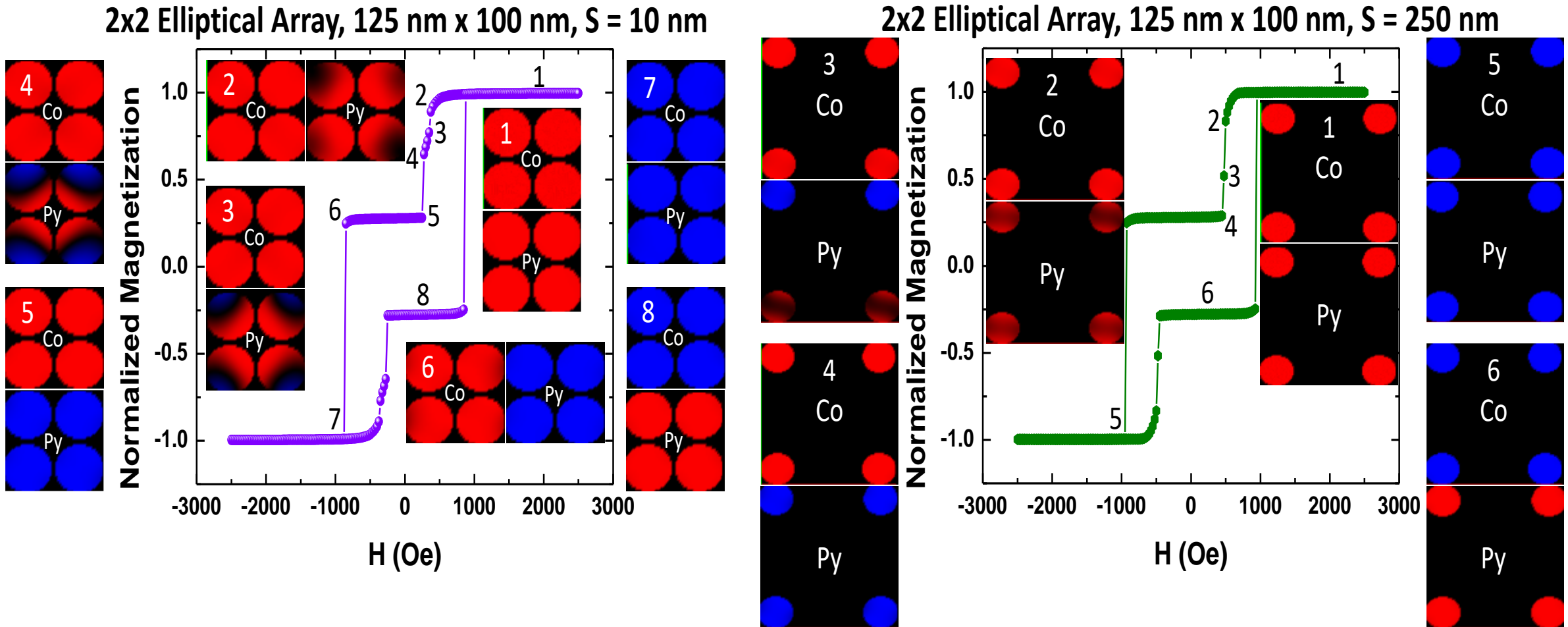

Fig. 7. Magnetization reversal states of Co and Py layers of the 2×2 arrays of elliptical spin valve elements with AR = 1.25, of two different cell spacings. The magnetic fields corresponding to which the magnetization maps are presented are marked by numbers.

We subsequently investigate how the magnetization reversal mechanism of the spin valve elements changes if we increase number of elements in a densely packed array. In Fig. 8. we show the magnetization maps of the Co and the Py layers at some important magnetic field values for a 3×3 elliptical array of spin valve elements with AR of 1.25 and a cell spacing of 10 nm. The magnetization maps are placed next to the corresponding hysteresis loop where the numbers indicate the field values at which the maps are simulated. At the +ve saturation field (1), both the Co and Py layers are aligned along the +x direction. At point (2), we notice that six of the Py elements lying on the left and right columns near the two boundaries form C-states, while the elements on the middle column retain its initial magnetization states. The Co elements also retain their original magnetization direction.  After the first switching a small plateau is formed (3) where the Py elements in the top and bottom rows fully switch their magnetization to -x direction, while two elements in the middle row remain in C-states and the central one retains its original magnetization. This is followed by another switching where a wider plateau is formed (4), where we notice that the Py layer elements have completely switched to -x magnetization direction. During the next switching another small plateau is formed (5) where three elements of the Co layer switch to -x magnetization direction while the remaining elements starts to show some edge nucleation regions for magnetization reversal.

This is followed by the final switching in the field sweep from +ve to -ve field direction (6), where both the layers have completely switched to -x magnetization direction

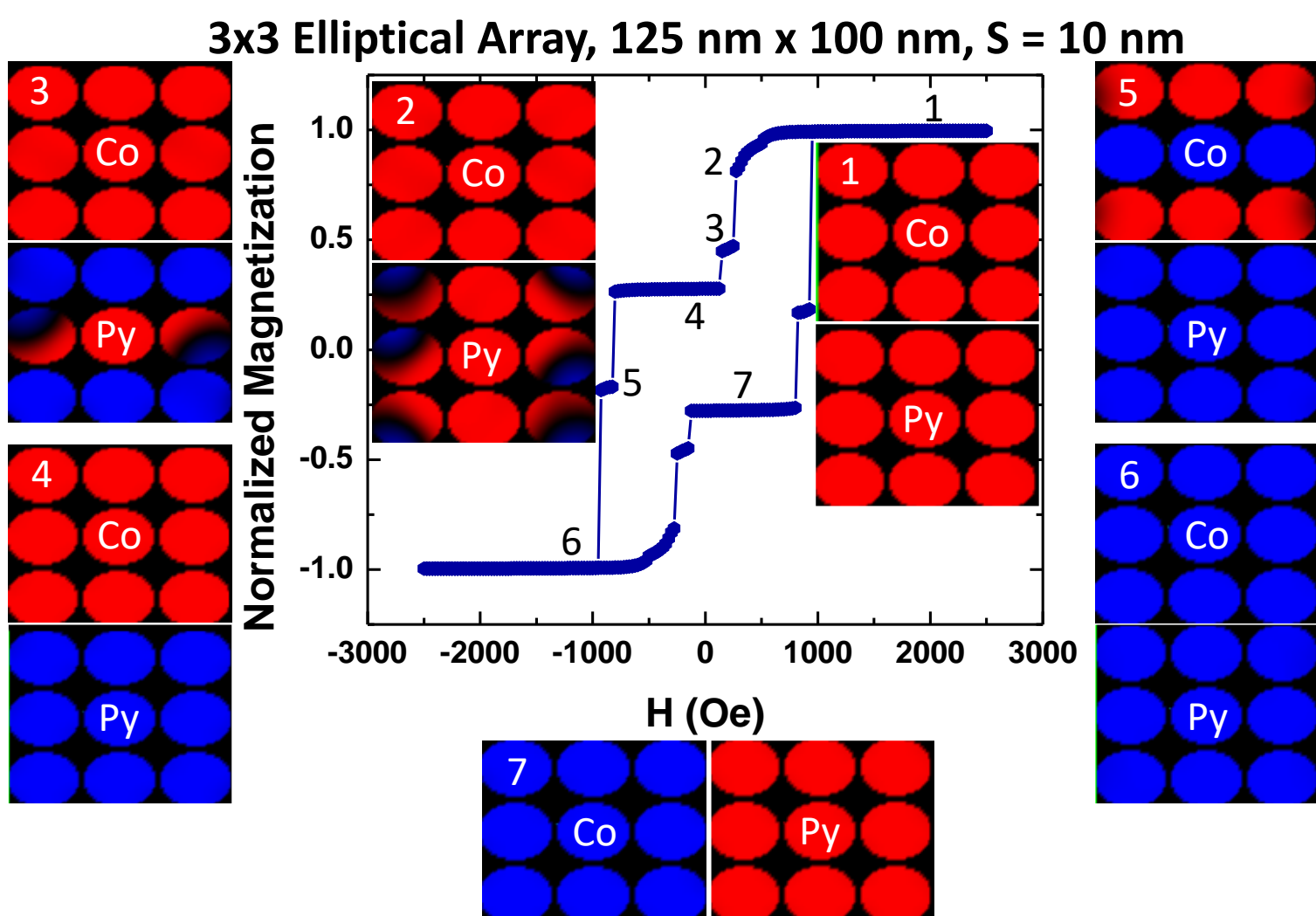

Fig. 8. Magnetization reversal states of Co and Py layers of the 3×3 arrays of elliptical spin valve elements with AR = 1.25, of cell spacing 10 nm. The magnetic fields corresponding to which the magnetization maps are presented are marked by numbers.

## 4. Conclusions

In summary, we have investigated the magnetic switching mechanism and magnetoresistive behavior in single spin valve elements of two different shapes and having a varying lateral aspect ratio and arrays of the elliptical elements having a constant aspect ratio but varying cell spacing by using computational micromagnetic simulations. We observe that the elements with higher AR show the $Ni_{80}Fe_{20}$ (Py) and Co layers forming antiparallel states in the plateau similar to synthetic antiferromagnets. As we reduce the AR, more complex quasi-uniform magnetic states are observed which are even more complicated for elliptical elements. The elliptical elements with the aspect ratio of 1.25 shows coherent and predictable switching behavior, showing its suitability for the application in magnetic memory elements. We observe a gradual increase in magnetoresistance (MR%) with the increase in AR of the spin valve elements, and the decrease in interelement spacing between the spin valve elements in their arrays. The array with the highest interelement spacing of 250 nm shows the most coherent and predictable switching behavior. Such complex magnetization reversal gives clear indication of the dominance of incoherent magnetization reversal in such nanoscale spin valve elements. The reversal mechanism is correlated with the various parameters of the magnetic hysteresis loops as well as the magnetoresistance of the individual elements as well as the array. While more studies will be required in the future on elements of different shapes, sizes, magnetic materials, interlayer interaction and geometry of the array, such correlations will set up guiding principles for the choice of the material, geometric shape and array geometry for optimal functionality of nanoscale spin valve arrays. Numerical simulations covering this hyper-parameter space and their optimization seems tedious. Here the usage of machine learning tools such as supervised machine learning including different regression methods, very fast simulated annealing algorithm, learning a mapping of spin configurations with simulation parameters etc. would be useful for the optimization process [22-23].

**Acknowledgements**

The author gratefully acknowledges TRC, S. N. Bose National Centre for Basic Sciences for funding and research fellowship and Dr. M. V. Kamalakar from Uppsala University, Sweden for useful discussions.

**Conflict of Interest**

The author declares no conflict of interest.